\documentclass[]{raa}            

\usepackage{graphicx,times}             
\usepackage{natbib}
\usepackage{amssymb,amsmath}
\usepackage{lineno}

\bibpunct{(}{)}{;}{a}{}{,}

\usepackage[a4paper=true,pagebackref=true]{hyperref}
\hypersetup{colorlinks = true, linkcolor = blue, anchorcolor = red, citecolor = blue, filecolor = red, pagecolor = red, urlcolor = red}
\usepackage{multirow} 

\begin{document}
   \title{Preliminary analysis on the noise characteristics of MWISP data
}

   \volnopage{Vol.0 (20xx) No.0, 000--000}      
   \setcounter{page}{1}          

   \author{Jia-Jun Cai
      \inst{1,2}
   \and Ji Yang
      \inst{2}
   \and Sheng Zheng
      \inst{1}
   \and Qing-Zeng Yan
      \inst{2}
   \and Shaobo Zhang
      \inst{2}
   \and Xin Zhou
      \inst{2}
   \and Haoran Feng
      \inst{2,3}
   }

   \institute{Center of Astronomy and Space Sciences, China Three Gorges University, Yichang 443002, China\\
        \and
             Purple Mountain Observatory and Key Laboratory of Radio Astronomy, Chinese Academy of Sciences, Nanjing 210033, China; {\it jiyang@pmo.ac.cn}\\
        \and
             School of Astronomy and Space Science, University of Science and Technology of China, Hefei 230026, China\\
\vs\no
   {\small Received~~20xx month day; accepted~~20xx~~month day}}
\abstract{ Noise is a significant part within a millimeter-wave molecular line datacube. Analyzing the noise improves our understanding of noise characteristics, and further contributes to scientific discoveries. We measure the noise level of a single datacube from MWISP and perform statistical analyses. We identified major factors which increase the noise level of a single datacube, including bad channels, edge effects, baseline distortion and line contamination. Cleaning algorithms are applied to remove or reduce these noise components. As a result, we obtained the cleaned datacube in which noise follows a positively skewed normal distribution. We further analyzed the noise structure distribution of a 3D mosaicked datacube in the range $l = 40^{\circ}_\cdot7$ to $43^{\circ}_\cdot3$ and $b = -2^{\circ}_\cdot3$ to 0$^{\circ}_\cdot3$ and found that noise in the final mosaicked datacube is mainly characterized by noise fluctuation among the cells.
}

   \authorrunning{J.-J Cai et al. }            
   \titlerunning{Analysis on Noise Characteristics of MWISP Data }  

   \maketitle

%
%
\section{Introduction}           
\label{sect:intro}

Three-dimensional (3D) datacubes are emerging widely in astronomy, especially for wide-range spectral surveys. Examples include the NRAO VLA Sky Survey (NVSS; \citealt{1998AJ....115.1693C}), the Red-Sequence Cluster Survey (RCS; \citealt{2005ApJS..157....1G}), The HI Nearby Galaxy Survey (THINGS; \citealt{2008AJ....136.2563W}), the Calar Alto Legacy Integral Field Area (CALIFA; \citealt{2013A&A...549A..87H}) survey, the ESO Very Large Telescope (VLT) with the Multi Unit Spectroscopic Explorer (MUSE) data products (\citealt{2014Msngr.157...13B}) and the HI4PI survey (\citealt{2016A&A...594A.116H}). Several Galactic CO surveys have also generated large-scale datacubes, such as the large scale CO survey of the Galactic center region (\citealt{1997A&AS..125...99B}), the FCRAO CO survey of the Outer Galaxy (\citealt{1998ApJS..115..241H}), the CO survey obtained with the CfA 1.2 m telescope (\citealt{2001ApJ...547..792D}), the Bell Laboratories $^{13}$CO survey (\citealt{2001ApJS..136..137L}) and the NANTEN CO survey (\citealt{2004ASPC..317...59M}). Such kind of datacubes is dominated by noise and frequently contaminated by artifacts.

Compared with one-dimensional (1D) spectral line data, the analysis of noise in a 3D datacube becomes more important for detecting weak signals in large-scale surveys. Statistics on the Galactic Ring Survey (GRS) has shown that a long tail in the histogram of root mean
square (rms) noise temperature represents the observation data obtained under various weather conditions, elevations, and observing modes (\citealt{2006ApJS..163..145J}). That study also discussed correlated noise among position-switching and On-The-Fly (OTF) mapping modes by analyzing the noise in 3D datacubes. A similar study on correlated noise is made for the Structure, Excitation and Dynamics of the Inner Galactic Interstellar Medium (SEDIGISM) survey (\citealt{2017A&A...601A.124S}). With the help of a noise histogram, the CO High-Resolution Survey (COHRS) revealed a faint residual tartan pattern due to the non-uniformity of the integration time (\citealt{2013ApJS..209....8D}). The study on the Mopra southern Galactic plane CO survey has revealed the images and probability distribution of the noise level (\citealt{2013PASA...30...44B}). A striped pattern overlaying the whole map has been discovered through exploring noise maps (\citealt{2013A&A...560A..24C}). The noise features are extracted in the research on $^{13}$CO/C$^{18}$O Heterodyne Inner Milky Way Plane Survey (\citealt{2016MNRAS.456.2885R}).

Understanding noise characteristics plays a pivotal role in developing source detection algorithms. Line Source Detection and Cataloguing (LSDCat), considering the noise property on very faint sources (\citealt{2017A&A...602A.111H}), is implemented on datacubes. Noise level variations are taken into account for detecting 3D sources using SOFIA (\citealt{2015MNRAS.448.1922S}). In source finding, parametrization, and classification for the extragalactic Effelsberg-Bonn HI Survey, a model to match the noise properties and correlated noise is involved (\citealt{2014A&A...569A.101F}). Correlated noise is also considered in a flexible noise model with fast and scalable methods (\citealt{2020A&A...638A..95D}). In blind detection of faint emission line galaxies in MUSE datacubes, the influence of noise is analyzed (\citealt{2020A&A...635A.194M}).

In this work, we analyzed noise of a typical cell datacube as well as a mosaicked datacube from Milky Way Imaging Scroll Painting (MWISP\footnote{http://www.radioast.nsdc.cn/mwisp.php}), a large-scale spectroscopic survey along the northern Galactic plane
in CO and its isotopes. Several attempts were applied to remove the extreme components in the noise and reduce the over-all noise level of the datacube. The single (cell) datacube and mosaicked datacube are analyzed in Section 3 and Section 4, respectively. Discussion and conclusion are provided in Section 5 and Section 6, respectively.


\section{Data description}
\label{sect:des}

\begin{figure}
	\centering
	\includegraphics[width=\textwidth, angle=0]{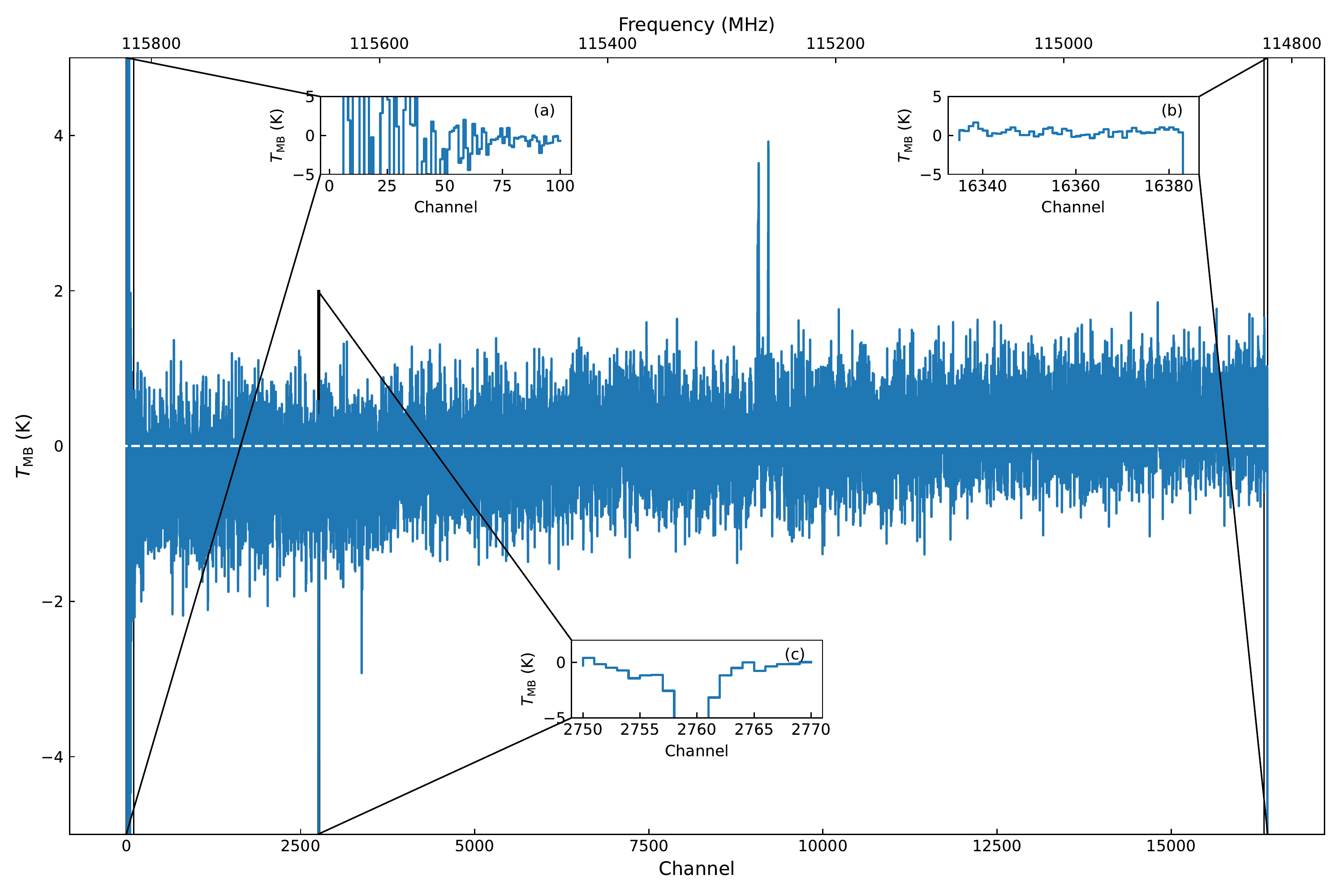}
	\caption{The spectral line at the coordinate (46, 46) on the $l-b$ plane of 0410$-$015U. Inset (a) shows the edge effect on the left side over about 75 channels. Inset (b) presents the edge effect on the right side containing several channels. Inset (c) displays three bad channels around the 2760th channel. The prominent features near the 9001th channel (0 km s$^{-1}$) of the spectral line illustrate the presence of CO line emission. The white dashed line delineates the baseline where $T\rm_{MB}$ = 0 K. }
	\label{Fig1}
\end{figure}

Datacubes in this study are available from the MWISP survey. A cell datacube, which is the basic unit of the data structure, is obtained by OTF (\citealt{2018AcASn..59....3S}) scanning along Galactic longitude ($l$) and latitude ($b$) over an area of 30$^{\prime}$ $\times$ 30$^{\prime}$ on the sky. A larger size of 45$^\prime_\cdot$5 $\times$ 45$^\prime_\cdot$5 is recorded during the re-gridding process, allowing an overlap between adjacent cells. Each cell datacube is 91 pixels by 91 pixels in $l-$ and $b-$ directions centered at coordinate (46, 46) with a pixel size of 30$^{\prime\prime}$ $\times$ 30$^{\prime\prime}$. The name of each cell datacube is denoted by LLLL$\pm$BBB, with the letter U at the end representing the upper sideband and the letters L and L2 at the end signifying the lower sideband. A datacube such as 0410$-$015U has 5996 spectral lines and the remaining 2285 positions are blank. Each cell datacube has a bandwidth of 1 GHz with a channel frequency interval of 61 kHz which corresponds to a velocity width of 0.16 km s$^{-1}$, providing 16 384 channels in the velocity (frequency) direction. The 9001th channel represents 0 km s$^{-1}$, near which $^{12}$CO $(J = 1\rightarrow0)$ emission at zero radial velocity appears (see Fig. \ref{Fig1}). The value of position-position-velocity (PPV) data denotes the main-beam brightness temperature in K. Mapping cells are mosaicked for a larger datacube. For more details about the MWISP survey, see \citet{2019ApJS..240....9S}.

To analyze the mosaicked datacube, we select 25 $^{12}$CO $(J = 1\rightarrow0)$ datacubes ($40^{\circ}_\cdot7 < l < 43^{\circ}_\cdot3$ and $-2^{\circ}_\cdot3 < b < 0^{\circ}_\cdot3$). Reading FITS (\citealt{2010A&A...524A..42P}) cubes is performed with the spectral-cube package\footnote{https://spectral-cube.readthedocs.io/en/latest/index.html} in Python language.

\section{Analysis of a single datacube}
\label{sect:analy}

We try to identify major noise components from different dimensions of the datacube. A 1D spectral line extracted from the coordinate (46, 46) on the $l-b$ plane of the 0410$-$015U datacube is depicted in Figure 1. Toward the left side of the spectral line in Figure 1, the frequency becomes higher where noise level is also higher due to atmospheric absorption. Insets (a) and (b) show the edge effects characterized by violent fluctuations on both ends of the spectral line. Inset (c) displays the bad channels near the 2760th channel like a spike embedded into the spectral line. The baseline of each spectrum was previously subtracted in a narrow velocity range, which may not be able to be used to derive a flat baseline over the whole velocity range. CO line emissions, bad channels, edge effects and the baseline distortion all contribute extra noise components to the statistics and these components need to be treated properly.

\subsection{Bad Channel}

Along the spectral axis ($v-$axis), bad channels occupy several channels with abnormally large amplitudes compared with their neighboring channels (see Fig. 1). In the previous study, bad channels, either with positive or negative temperature values, usually caused by poor channel performance are recognized as a part of excessive noises (e.g., \citealt{2001JKAS...34....1L}). Inspection of each line profile confirms that some spectral lines may not have bad channel spikes, but some other spectral lines have more than one spike. Figure 2 displays the integrated image from the 2750--2770th channel containing bad channels. Stripes of varying intensities in Figure 2 result from bad channels with varying amplitudes. The distribution of stripes is in accordance with the scanning directions.

In a 3D datacube, bad channels contribute only a small part of the total voxels. However, they introduce prominent noise over a localized area in the datacube and therefore should be carefully removed. With the aid of the integrated images, we manually identify all the prominent bad channels within the whole datacube. Then, we flagged out these bad channels. These bad channels will not contribute to the noise statistics hereafter. By checking spectral lines and integrated images, we found bad channels are mainly concentrated within the 2750--2770th, 3370--3390th and 4900--4920th channels. The features of bad channels can be summarized as follows: abnormally large amplitudes, an abruptly increasing contour, and the distribution along the scanning directions. According to these features, bad channels can be automatically removed by pipelines.




\begin{figure}
	\centering
	\includegraphics[width=\textwidth, angle=0]{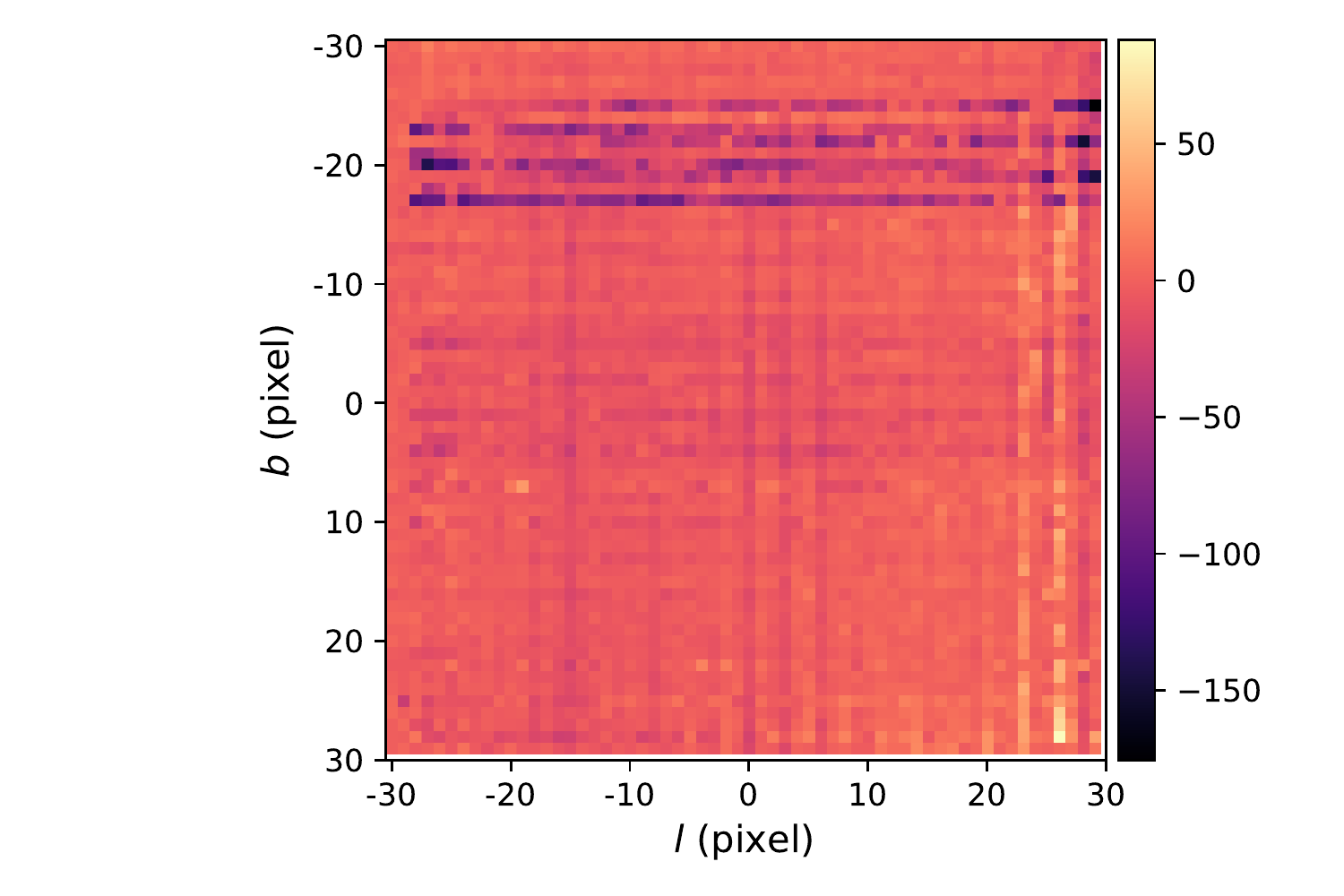}
	\caption{An intensity image integrated over 2750--2770th channels of 0410$-$015U datacube. To highlight the bad channels more clearly, we restrict the range of $l-b$ in 30 pixels$\times$30 pixels. The image is in units of K km s$^{-1}$. }
	\label{Fig2}
\end{figure}

\subsection{Edge Effect}

The edge effects in a datacube exist in three dimensions and they can be classified into two categories based on their underlying causes. The edge effects in the $v-$direction are reflected in the abnormal fluctuations at both ends of each spectral line, resulting from non-linearity of the sideband edges. The difference in values between adjacent channels at both ends of the spectral line can reach several hundreds, and the edge effects at the left end generally occupy more channels than those at the right end.

The median absolute deviation (henceforth MAD) method (\citealt{doi:10.1080/01621459.1974.10482962}) is utilized to identify the edge effects, which is a robust method (\citealt{ROTA1986123}). MAD is defined as follows
\begin{equation}
	{\rm{MAD}} = b\ {\rm{med}}\left|x_i - {\rm{med}}(x_i)\right|,
	\label{MAD}
\end{equation}
where $x_i$ is the $n$ original data and ${\rm{med}}$ is the median of the data set. The constant $b$ is usually set to $b = 1.4826$ under the assumption of normality of the data set. In the region with edge effects, the fluctuations are abnormally large where data values can be regarded as outliers. The locations of these outliers can be used to determine the regions where edge effects exist. By computing
\begin{equation}
	\frac{\left|{x_i}\ -\ {\rm{med}}(x_i)\right|}{\rm{MAD}}
	\label{criterion}
\end{equation}
for each $x_i$, those $x_i$ for which this formula exceeds a certain threshold (here we use 3) are flagged.

The red dots in panels (a) and (b) of Figure 3 are outliers as ascertained by the MAD method, and the channels where they are located can be considered as the range of edge effects in the $v-$direction. After examining 25 datacubes, the first 100 channels and the last 10 channels are identified as edge effects in the $v$ direction that must be removed.

Edge effects in $l-$ and $b-$ directions could be introduced when a multi-beam receiver is scanning along the Galactic longitude and/or latitude (\citealt{2006ApJS..163..145J}; \citealt{2018AcASn..59....3S}). Due to the insufficient exposure compared with the central 30$'\times$30$'$ (umbra), the amplitudes of noise around the edges (penumbra) are quite large, typically between 5$\sim$10 K. We identified the edge effects in the $l-$ and $b-$ directions applying the MAD method for ranges illustrated in panel (c) of Figure 3. Since edge effects in $l-$ and $b-$ directions are due to the decrease of exposure, and is almost compensated by the overlap of neighboring scans during the mosaic process for a large survey map (see Sect. 4.1 below), therefore we do not introduce any operation here.

    \begin{figure}
	\centering
	\includegraphics[width=\textwidth, angle=0]{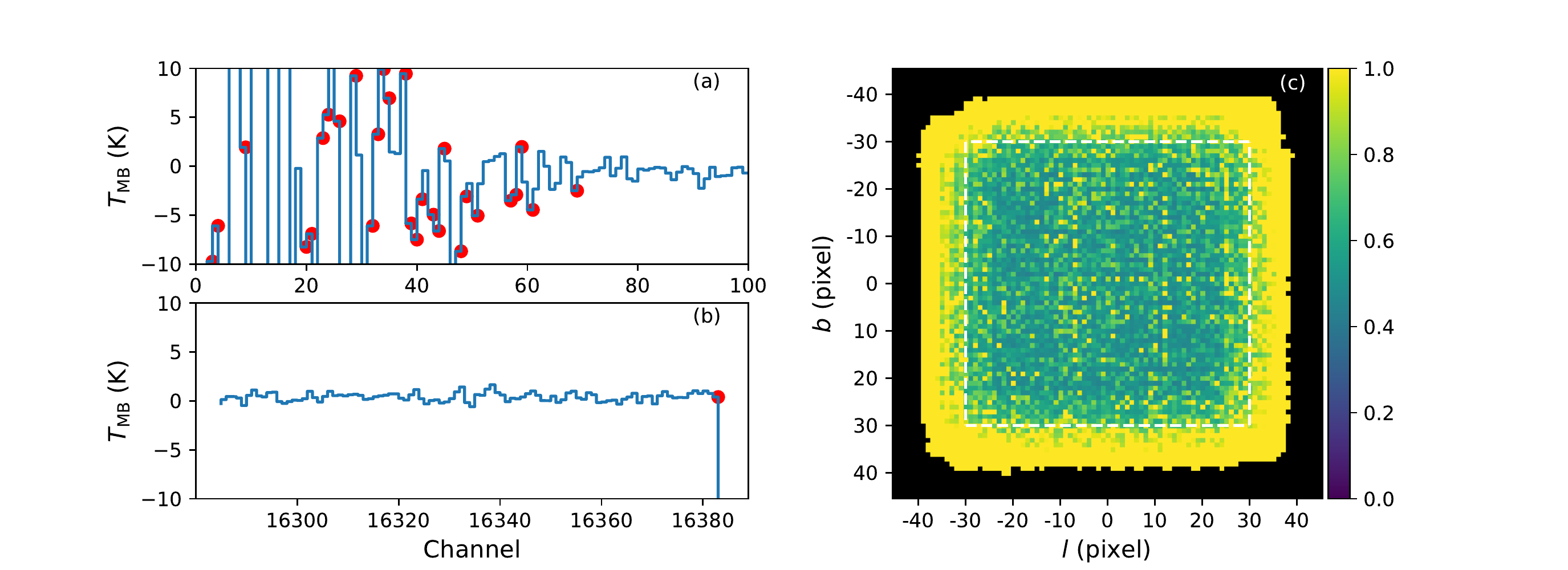}
	\caption{Edge effects in three directions of the datacube 0410$-$015U. Panels (a) and (b) show the edge effects at both ends of the central spectral line in the $v-$direction. The red dots represent the range of edge effects in the $v-$direction marked by the MAD method. Panel (c) features the image of rms noise, in units of K, calculated using all the channels within $-$100 km s$^{-1}$ to 100 km s$^{-1}$. The colorbar is limited by $\le$2 K in order to display the edge effects clearly. }
	\label{Fig3}
\end{figure}

\subsection{Baseline Inclination}

\begin{figure}
	\centering
	\includegraphics[width=\textwidth, angle=0]{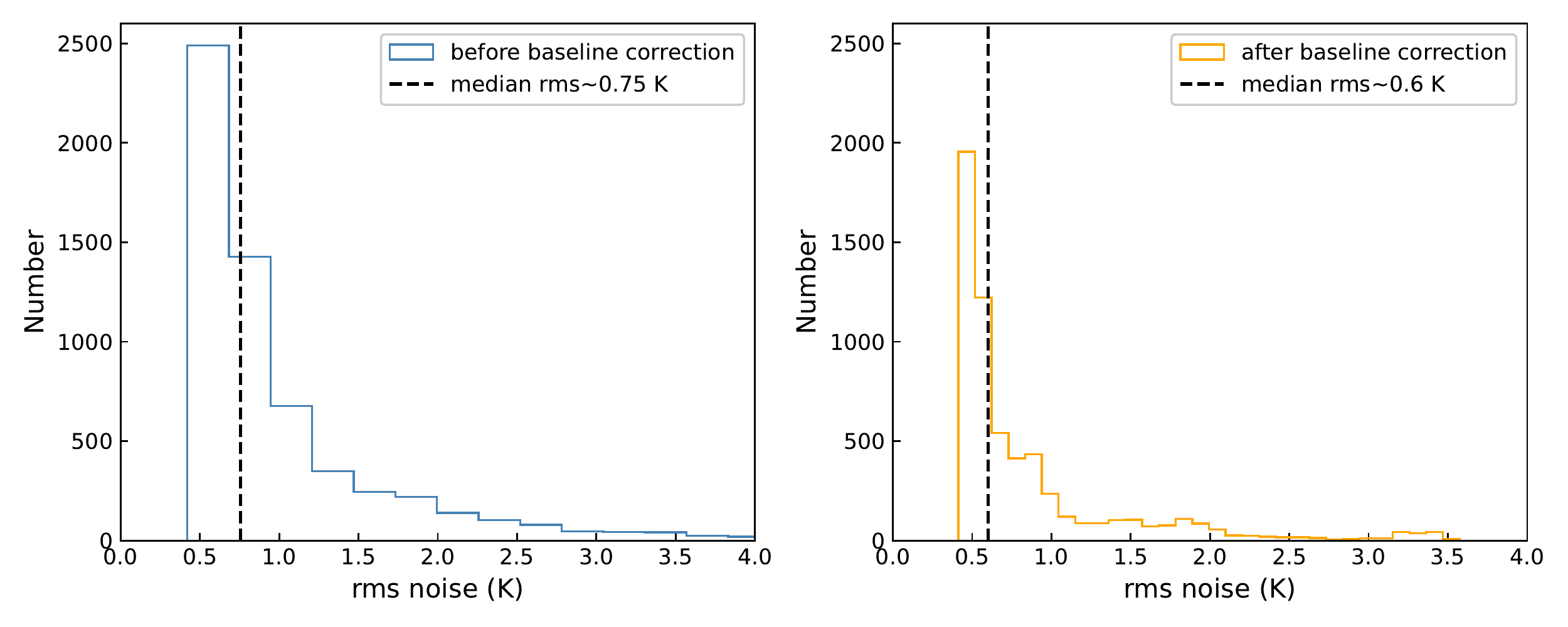}
	\caption{The comparison of noise distribution before and after the baseline correction. The dashed line correspond to the median rms noise of the distribution. The rms noise is limited by $\le$4 K for comparison.}
	\label{Fig4}
\end{figure}

We manually examined all the 5996 spectral lines within the 0410$-$015U datacube and found that most of the original baselines are inclined, with their slope in the range $-$1.32$\times$10$^{-3}$ to 1.59$\times$10$^{-3}$. The bright spots in the panel (c) of Figure 3 are in fact caused by the poor baselines. To further improve the baseline quality, a linear least squares method was applied to perform a first-order baseline fit to these spectral lines and we subtracted fitted baselines from the original data for later noise statistics. The improvement of baseline, measured in terms of the rms noise level, is from 0.75 K to 0.6 K (see Fig. \ref{Fig4}).

\subsection{Influence by CO line emission}

The presence of CO line emission introduces extra counts to the noise statistics. We examine the influence of line emission by using the datacube after baseline subtraction.  Figure 5(a) plots the noise amplitude calculated by rolling statistics along the $v-$axis with a width of 81 channels. The in-band noise varies significantly. A gradual increase of noise towards high frequency is due to the sharp increase of atmospheric opacity in the upper sideband where the  $^{12}$CO $(J = 1\rightarrow0)$ line is located. A prominent feature between two dashed lines within the velocity range from $-$100 km s$^{-1}$ to 100 km s$^{-1}$ in panel (a) is due to the presence of CO emission lines. As long as such an emission line exists, it will bring additional components into the noise statistics.

\begin{figure}
	\centering
	\includegraphics[width=\textwidth, angle=0]{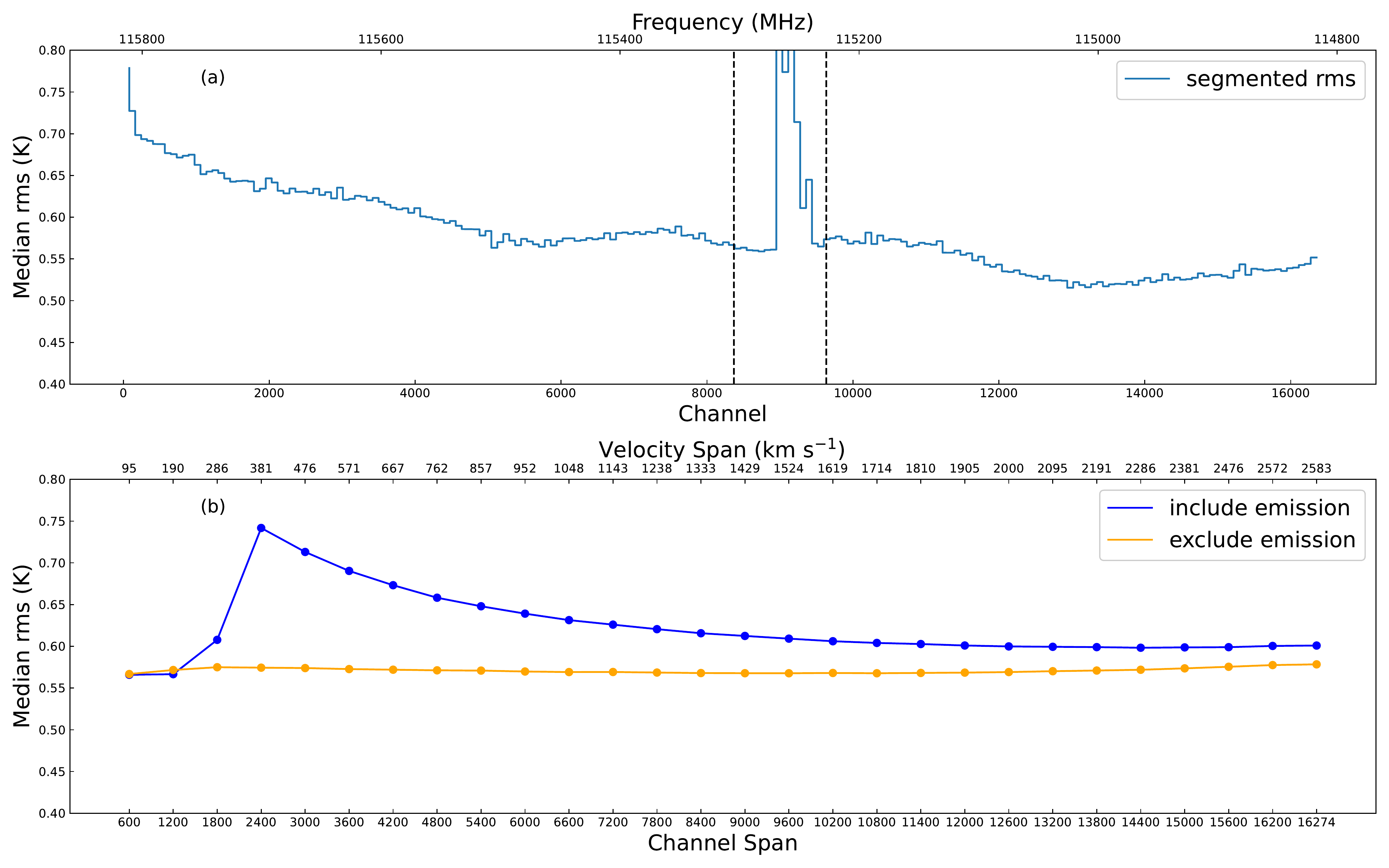}
	\caption{Rolling statistics along the band. The panel (a) plots the rolling statistics calculated in each slice including 81 channels on spectral lines in the datacube 0410$-$015U. Panel (b) depicts the influence of the emission line by gradually expanding channel spans centered at the line. The blue line signifies the case including the line and yellow is for excluding the line by flagging out the channels within  $-$100 km s$^{-1}$ to 100 km s$^{-1}$. }
	\label{Fig5}
\end{figure}

To further illustrate the impact of emission lines on the noise statistics, rolling statistics of the slices with increasing spans are calculated with the reference position at the line center and a step of 600 channels, as displayed in Figure 5(b). For the blue line in panel (b), no emission is included in the span at the start. Once the line emission is contained as the span expands, the rms drastically increases. As long as the number of channels increases, the median rms decreases gradually but remains prominent. Therefore, statistics on datacube noise level should consider any possible contamination by line emissions.

\begin{table}[htbp]
	\centering
	\caption{Extreme Noise Components and Treatment}
	\resizebox{\textwidth}{!}{
		\begin{tabular}{c|c|c|c}
			\hline
			\hline
			Component & Position & Treatment & Decrease of rms after treating \\
			\hline
			Bad channel & several channels in $v-$axis & flag out & 0.10 K \\
			Edge effect in $v-$direction & both ends of the spectral line & remove & 10.14 K \\
			Baseline inclination & the whole spectral line & subtract & 0.15 K \\
			Line contamination & several windows in $v-$axis & flag out & 0.02--0.27 K \\
			\hline
		\end{tabular}%
		\label{tab:factors_property}%
	}
\end{table}%

Table 1 lists the extreme noise components and corresponding operations for a cell datacube. In practical use, these operations may be realized by an advanced pipeline.


\section{Analysis of a Mosaicked Datacube}

\subsection{Non-uniformity among Datacubes in a Mosaic}
\begin{figure}
	\centering
	\includegraphics[width=\textwidth, angle=0]{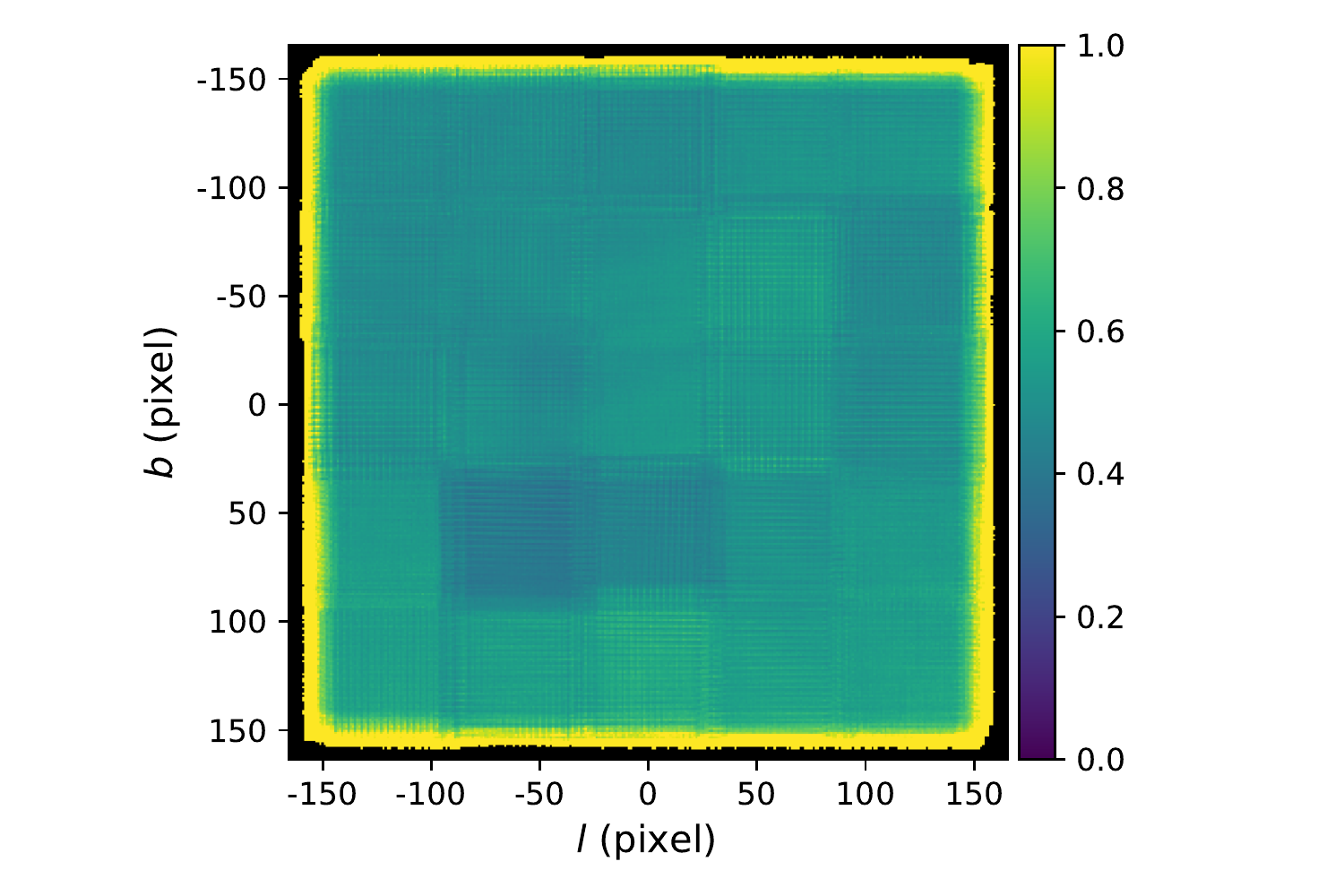}
	\caption{1D noise image of mosaicked datacube after cleaning. The 1D rms noise was calculated along the whole spectral axis. The scattering of noise level among cell units is visible. This noise image is in units of K. }
	\label{Fig6}
\end{figure}

Twenty-five cell datacubes are mosaicked into a single datacube with a size of 331 pixels$\times$329 pixels$\times$16384 channels. Each row has 331 pixels and each column has 329 pixels. Figure \ref{Fig6} shows the image of rms noise calculated using all remaining available channels at each pixel for the mosaicked datacube. A smooth overlap between an adjacent datacube was achieved by the designed mosaicking, as can be judged from the noise image, confirming that the scanning size and overlap arrangements between neighboring cell datacubes have been reasonablly designed.

In order to examine rms noise fluctuations among individual cells, we plot in Figure 7 the rms distributions cut along the center of noise image in $l-$ and $b-$ directions. The gray color marks the overlapped regions. The rms values are in fact fluctuating, mainly due to the mutual scattering among individual cell datacubes. A search from Table A.1 confirms that the median rms ranges from 0.49 to 0.72 K.

\begin{figure}
	\centering
	\includegraphics[width=\textwidth, angle=0]{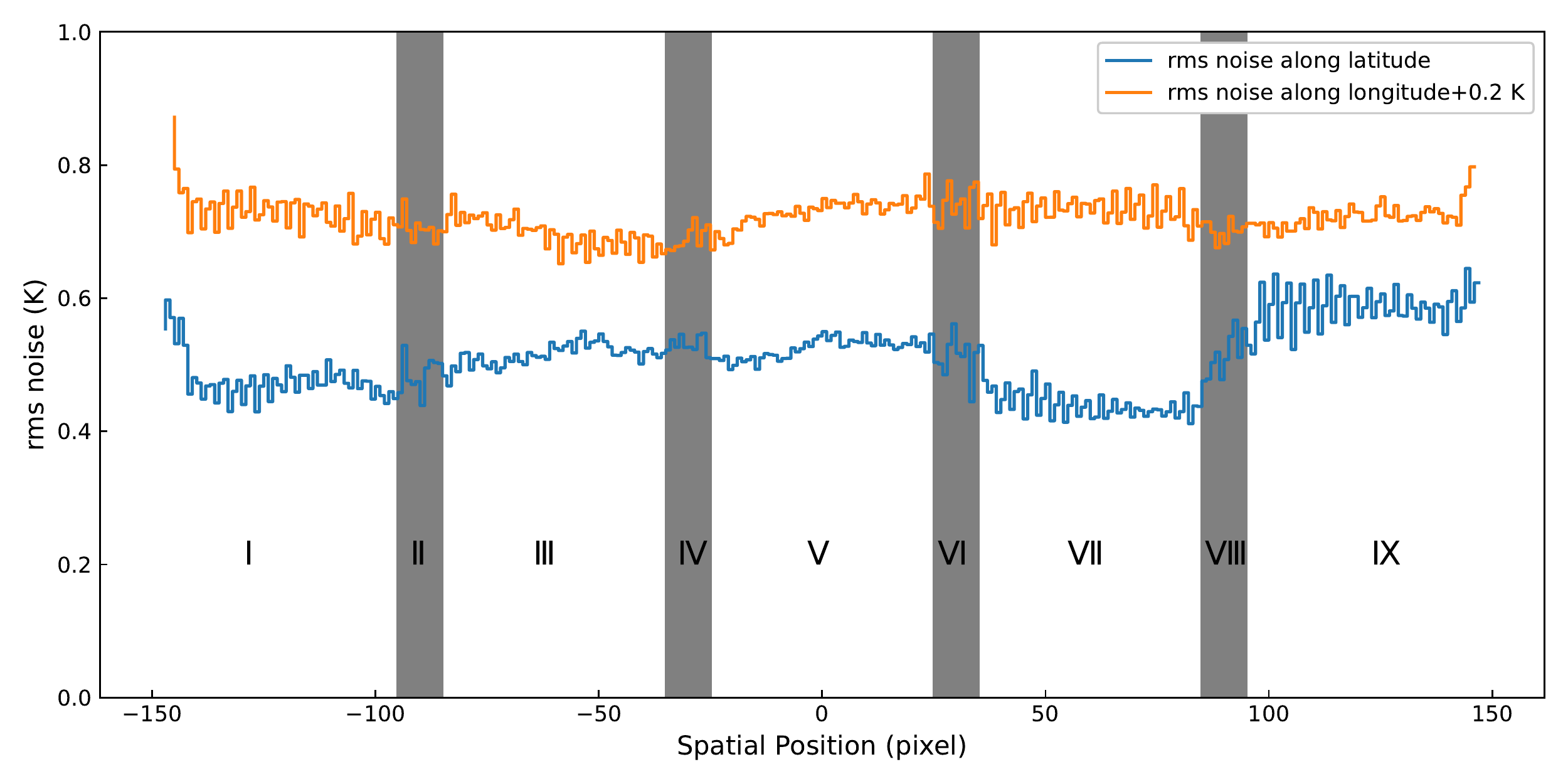}
	\caption{Rms plot along the Galactic longitude and latitude through the center of the noise image for a mosaicked datacube. For clarity, an offset by 0.2 K was added to the longitude plot. The vertical gray bars illustrate the overlapped regions during the mosaic process.}
	\label{Fig7}
\end{figure}

Table A.1 provides statistics for the nine individual cell datacubes in Figure 6. A striking feature is that the median rms of cell datacube \uppercase\expandafter{\romannumeral9} along latitude is 34.1$\%$ higher than that of cell datacube \uppercase\expandafter{\romannumeral7}. On the other hand, the standard deviations of these cell datacubes are small, but in contrast, the fluctuation among the cell datacubes dominates the final mosaicked datacube. This result indicates that the difference in noise levels between cell datacubes plays a major role in the inhomogeneous noise image for the final image.

We note that either before or after the cleaning procedure, the statistical distribution of noise in the mosaicked datacube is similar to those in single datacubes in the sense of the major Gaussian component, as affirmed in Figure 8. The noise distributions of both single and mosaicked datacubes have prominent tails. This tail was contributed mainly by those from the penumbra part where the effective integration is less than the central umbra part. Such effect is clearly demonstrated by the inset in Figure 8.

\subsection{Three Dimensional Noise Statistics}

Table 2 lists the statistics of the mosaicked datacube for cases either before and after cleaning, along with those for individual cell datacubes. The 1D rms noise was calculated along the $v$ axis using all remaining channels excluding line emissions. The data shows that significant improvement of noise properties has been achieved.

\begin{figure}
	\centering
	\includegraphics[width=\textwidth, angle=0]{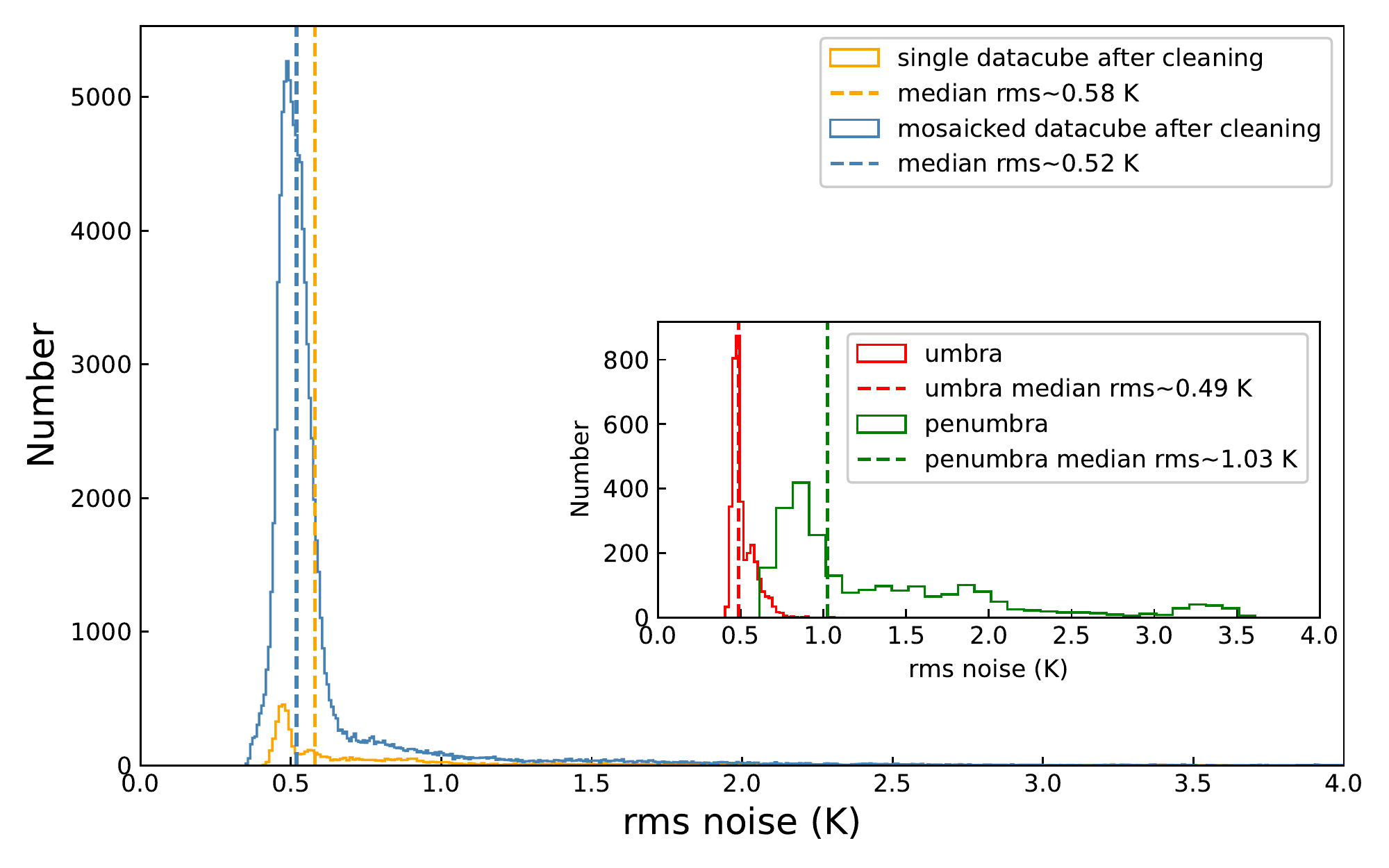}
	\caption{A comparison of noise distribution between a single cell datacube and mosaicked datacube after cleaning. The dashed lines correspond to the median rms noise of the distributions. The rms noise is limited by $\le$4 K in order to display the noise distribution clearly.}
	\label{Fig8}
\end{figure}

In principle, any calculation of noise should depend on the definition of a statistical neighborhood. For a 3D PPV datacube, we attempt to build the 3D neighborhood and calculate a 3D rms noise. We construct a series of cubic neighborhoods with the reference position at the central spectral line of the mosaicked datacube. The initial box size is 2 pixels$\times$2 pixels$\times$10 channels, then we expand it in three dimensions in constant steps until any dimension can no longer be expanded. The step along the $l-$ and $b-$ directions is 2 pixels, and the step along the $v-$direction is 10 channels. The results are depicted in Figure 9. Two features can be found: (1) The 3D noise distributions corresponding to different datacubes have similar trends. The noise variation between $1\times10^7$ to $4\times10^7$ voxels is relatively slow, especially when the noise is calculated at a lower frequency on the spectral line. At about $4\times10^7$ voxels, the noise begins to increase. (2) The 3D noise amplitude systematically changes due to the in-band noise variation along the frequency (velocity) axis, which is consistent with the 1D rolling statistics displayed in Figure 4(a).

\begin{table}[htbp]
	\centering
	\caption{The comparison of noise statistics before and after cleaning extreme noise components in single and mosaicked datacubes. The rms noise is calculated toward the $v-$direction using all remaining channels excluding emissions.}
	\resizebox{\textwidth}{!}{
		\begin{tabular}{c|c|c|c|c}
			\hline
			\hline
			Noise statistics & Original single datacube & Single datacube after cleaning & Original mosaicked datacube & Mosaicked datacube after cleaning \\
			\hline
			Min. (K) & 1.16  & 0.41  & 0.63  & 0.35 \\
			Max. (K) & 623.04 & 3.60  & 400.02 & 7.83 \\
			Mean (K) & 16.48 & 0.85  & 9.22 & 0.61 \\
			Median (K) & 10.98 & 0.58  & 9.73 & 0.52 \\
			Std. (K) & 30.60 & 0.61  & 5.09  & 0.37 \\
			\hline
		\end{tabular}%
		\label{tab:noise_property}%
	}
\end{table}%

Two dashed lines indicate the number of voxels contained in the smallest and largest molecular cloud found in the first Galactic quadrant using the DBSCAN\footnote{https://scikit-learn.org/stable/modules/generated/sklearn.cluster.DBSCAN.html} algorithm (see full details in \citealt{2020ApJ...898...80Y}), respectively. In this way, we can estimate the 3D noise level near any molecular cloud.

\begin{figure}
	\centering
	\includegraphics[width=\textwidth, angle=0]{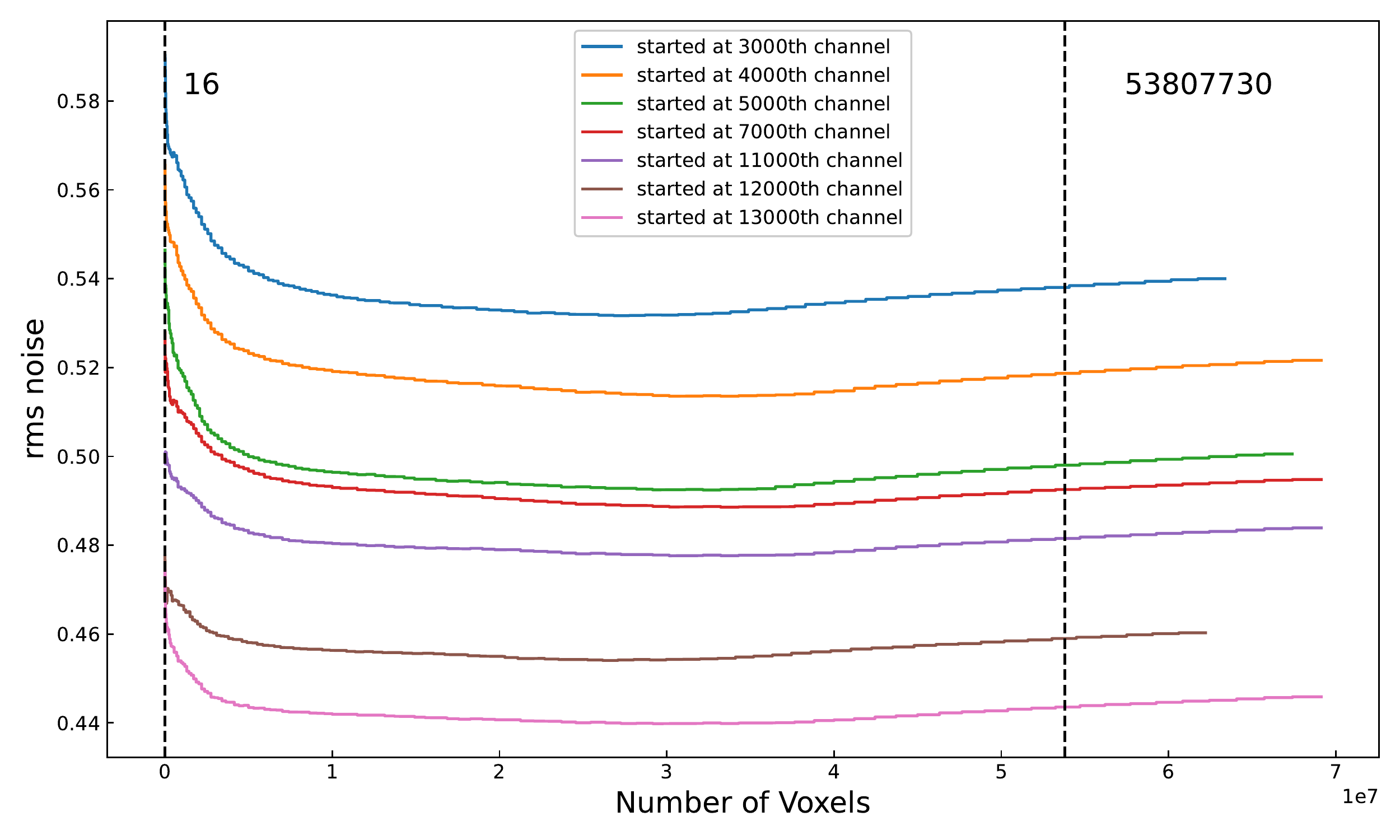}
	\caption{3D rms noise in PPV space versus the 3D voxels (volume).  Lines of different colors indicate different starting positions at the $v-$axis. Two dashed lines outline the voxel  range of CO cloud samples identified by the DBSCAN algorithm. }
	\label{Fig9}
\end{figure}

\section{Discussion}
\label{sect:discussion}

To analyze the noise, the first step is to clean the extra noise from the data. As already described in this study, the major factors that need to be cleaned for MWISP datacubes are mainly bad channels, edge effects, baseline distortion, and line components. Among these effects, the edge effect increases the noise level the most.

The noise level decreases significantly after cleaning these factors, and the resultant datacube contains the generic noise floor which can be a good measure of the detection limit. 

As has been shown by the analysis, the spectral noise level within a datacube is not uniform in the $l-$, $b-$ and $v-$ directions. The uneven noise level in the $v-$direction is mainly caused by a sharp increase of atmospheric opacity in the upper sideband due to the atmospheric absorption. The uneven noise level on the $l-b$ plane stems mainly from the difference between the datacubes obtained under different observation conditions. 
An improvement can be implemented in a future survey to a more adaptive integration, e.g., by developing a dynamic noise level monitoring and management system (e.g., \citealt{2001ApJ...547..792D}) to  ensure that the noise level between different mapping cells is within a certain range.

It may be more appropriate to describe noise characteristics in terms of distributions rather than constant values. Several studies have tried to estimate the noise distribution of a 3D datacube (\citealt{2020ApJ...897..122L}; \citealt{2021MNRAS.502.1218R}). Further work is required to establish the noise distribution model suitable for MWISP data, so as to detect faint sources.

So far, 1D spectral noise level has been regarded as a measure of the PPV datacube. When there are only 1D spectral line data and the velocity range of the spectral line is relatively narrow, it is reasonable to use the baseline rms of the spectral line neighborhood to represent the rms of this spectral line. However, when the velocity range of spectral components is large, or the spatial  scales of clouds are large, 1D noise cannot represent the noise distribution of a 3D datacube with enough accuracy. 

In these cases, it is necessary to put forward a more general definition of noise in a 3D datacube. Specifically, it is recommended that, for a measurement of parameters for clouds of different cloud sizes (d$l-$d$b-$d$v$) at a certain location ($l$, $b$, $v$), a noise value at the corresponding location and scale should be adopted. Further analysis on the spatial dependence of noise is helpful to characterize the noise properties of the entire datacube in PPV space, and based on such efforts, more accurate measurements of large-scale CO emission can be expected.

\section{Conclusions}
\label{sect:conclusion}
We made a detailed statistical analysis on the noise property for 25 datacubes of the $^{12}$CO $(J = 1\rightarrow0)$ survey in the range $l = 40^{\circ}_\cdot7$ to $43^{\circ}_\cdot3$ and $b = -2^{\circ}_\cdot3$ to 0$^{\circ}_\cdot3$. Our main conclusions are as follows:

1. We identified the major components of the extra noise, including edge effect, baseline distortion, bad channel and line contamination. After appropriate treatment,the remaining noise follows a positively skewed normal distribution.

2.  Within a 3D datacube, large-scale gradient is a significant factor which cannot be ignored during signal detection and measurements. More elaborat treatment of 3D noise should be introduced for the case of a datacube with an inhomogeneous noise distribution.

3.  For a large-scale mosaicked mapping, the fluctuation of noise level in different mapping cells is a fundamental issue in characterizing the final rms noise level. Better control of uniformity of integration time among mapping cells is essential to the final large-scale mosaicked image.

\begin{acknowledgements}
We would like to show our gratitude to support members of the MWISP group, Xin Zhou, Zhiwei Chen, Shaobo Zhang, Min Wang, Jixian Sun, and Dengrong Lu, and observation assistants at PMO Qinghai station for their longterm observation efforts. We are also grateful to Yang Su, Yan Sun, Chen Wang and Lixia Yuan for their useful discussions. We appreciate the referee's helpful comments. The MWISP project is supported by the National Key R\&D Program of China (2017YFA0402701) and Key Research Program of Frontier Sciences of CAS (QYZDJ-SSW-SLH047). This work is partially supported by the National Natural Science Foundation of China (Grant No. U2031202). This study has made of the Astrophysics Data System operated by the Smithsonian Astrophysical Observatory under NASA Cooperative Agreement.

\end{acknowledgements}

\appendix                  

\section{25 $^{12}$CO $(J = 1\rightarrow0)$ datacubes noise statistics figures and tables}
We present full-channel noise histograms of 25 datacubes and rms noise level in Figure A.1 and Table A.1, respectively.
\clearpage 
    \begin{figure}
	\centering
	\includegraphics[width=\textwidth, angle=0]{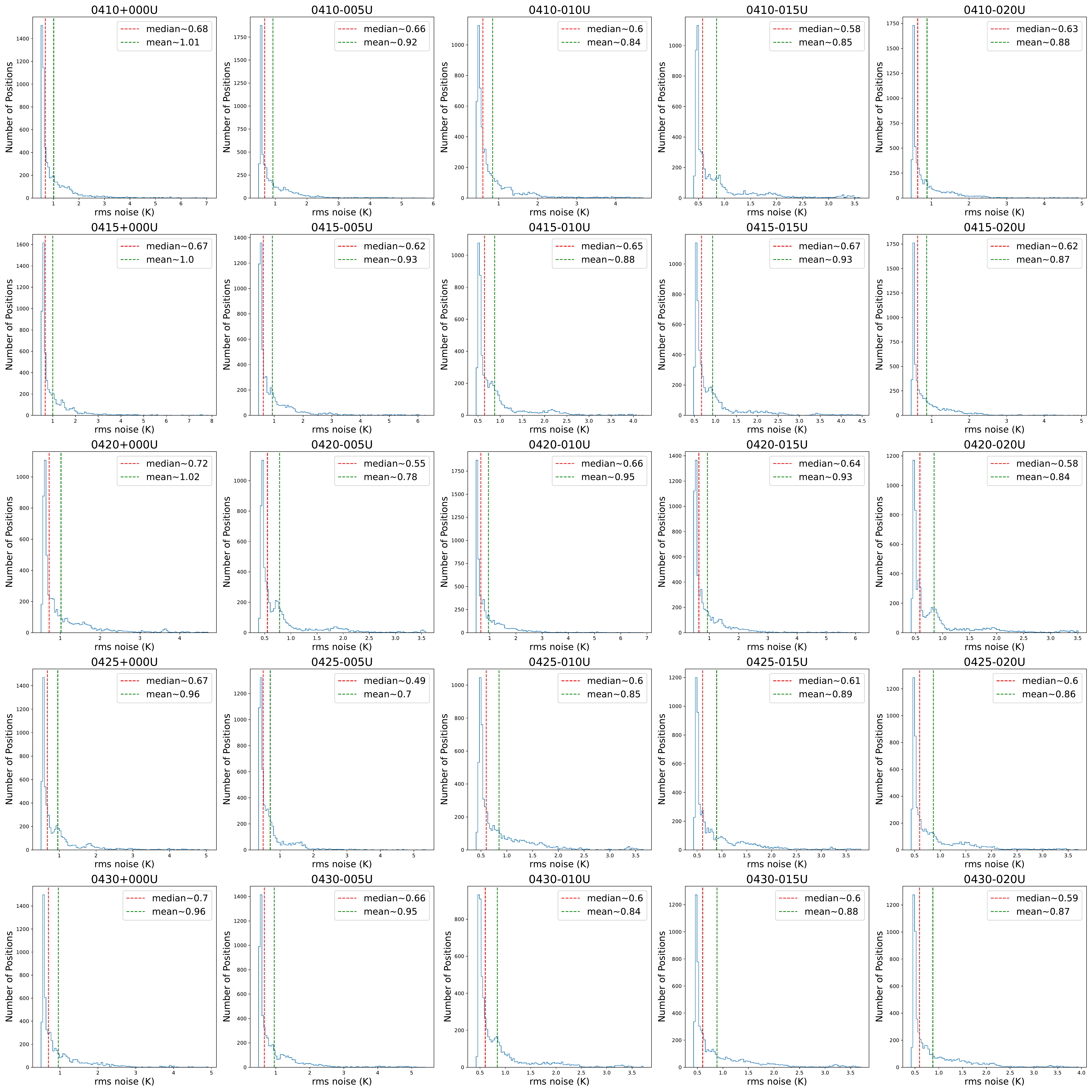}
	\caption{Full-channel noise histograms for 25 datacubes. }
	\label{FigA1}
\end{figure}

\begin{table}[htbp]
	\centering
	\caption{rms level for 25 datacubes}
	\resizebox{\textwidth}{!}{%
	\begin{tabular}{c|ccccc}
		\hline
		\hline
		\multirow{2}[2]{*}{Datacube} & \multicolumn{5}{c}{rms noise level} \\
		& Min. (K) & Max. (K) & Mean (K) & Median (K) & Std. (K) \\
		\hline
		0410$+$000U & 0.51  & 7.06  & 1.01  & 0.68  & 0.77 \\
		0410$-$005U & 0.47  & 5.76  & 0.92  & 0.66  & 0.63 \\
		0410$-$010U & 0.42  & 4.70  & 0.84  & 0.60  & 0.62 \\
		0410$-$015U & 0.41  & 3.61  & 0.85  & 0.58  & 0.61 \\
		0410$-$020U & 0.46  & 4.90  & 0.88  & 0.63  & 0.58 \\
		0415$+$000U & 0.49  & 7.83  & 1.00  & 0.67  & 0.79 \\
		0415$-$005U & 0.45  & 6.26  & 0.93  & 0.62  & 0.77 \\
		0415$-$010U & 0.46  & 4.22  & 0.88  & 0.65  & 0.57 \\
		0415$-$015U & 0.48  & 4.46  & 0.93  & 0.67  & 0.65 \\
		0415$-$020U & 0.45  & 4.91  & 0.87  & 0.62  & 0.56 \\
		0420$+$000U & 0.51  & 4.71  & 1.02  & 0.72  & 0.68 \\
		0420$-$005U & 0.38  & 3.58  & 0.78  & 0.55  & 0.56 \\
		0420$-$010U & 0.48  & 6.86  & 0.95  & 0.66  & 0.72 \\
		0420$-$015U & 0.46  & 6.18  & 0.93  & 0.64  & 0.72 \\
		0420$-$020U & 0.42  & 3.51  & 0.84  & 0.58  & 0.60\\
		0425$+$000U & 0.50  & 5.05  & 0.96  & 0.67  & 0.69 \\
		0425$-$005U & 0.35  & 5.35  & 0.70  & 0.49  & 0.52 \\
		0425$-$010U & 0.40  & 3.65  & 0.85  & 0.60  & 0.56 \\
		0425$-$015U & 0.43  & 3.79  & 0.89  & 0.61  & 0.62 \\
		0425$-$020U & 0.43  & 3.69  & 0.86  & 0.60  & 0.58 \\
		0430$+$000U & 0.50  & 4.91  & 0.96  & 0.70  & 0.61 \\
		0430$-$005U & 0.49  & 5.41  & 0.95  & 0.66  & 0.71 \\
		0430$-$010U & 0.42  & 3.72  & 0.84  & 0.60  & 0.56 \\
		0430$-$015U & 0.43  & 3.70  & 0.88  & 0.60  & 0.61 \\
		0430$-$020U & 0.42  & 3.93  & 0.87  & 0.59  & 0.60\\
		\hline
	\end{tabular}%
	\label{25datacubes_rms_level}%
}
\end{table}%



\clearpage 
\bibliographystyle{raa}
\bibliography{msRAA-2021-0229.R1}

\label{lastpage}

\end{document}